\documentclass[twocolumn,english,amsart,showpacs,preprintnumbers,amsmath,amssymb,floatfix]{revtex4-1}
\usepackage{tikz,xcolor}
\usepackage{amsmath}
\usepackage[colorlinks = true,
linkcolor = blue,
urlcolor  = blue,
citecolor = blue,
anchorcolor = blue]{hyperref}

\definecolor{lime}{HTML}{A6CE39}
\DeclareRobustCommand{\orcidicon}{%
	\begin{tikzpicture}
		\draw[lime, fill=lime] (0,0)
		circle [radius=0.16]
		node[white] {{\fontfamily{qag}\selectfont \tiny ID}};
		\draw[white, fill=white] (-0.0625,0.095)
		circle [radius=0.007];
	\end{tikzpicture}
	\hspace{-2mm}
}

\foreach \x in {A, ..., Z}{%
	\expandafter\xdef\csname orcid\x\endcsname{\noexpand\href{https://orcid.org/\csname orcidauthor\x\endcsname}{\noexpand\orcidicon}}
}

\usepackage[T1]{fontenc}
\usepackage[latin9]{inputenc}
\usepackage{color}
\usepackage{array}
\usepackage{amstext}
\usepackage{graphicx}
\usepackage{esint}
\usepackage{rotating}
\usepackage{appendix}
\usepackage{float}

\usepackage[font=small,labelfont=bf]{caption}
\usepackage{xcolor}
\usepackage{ulem}

\makeatletter

\@ifundefined{textcolor}{}
{%
	\definecolor{BLACK}{gray}{0}
	\definecolor{WHITE}{gray}{1}
	\definecolor{RED}{rgb}{1,0,0}
	\definecolor{GREEN}{rgb}{0,1,0}
	\definecolor{BLUE}{rgb}{0,0,1}
	\definecolor{CYAN}{cmyk}{1,0,0,0}
	\definecolor{MAGENTA}{cmyk}{0,1,0,0}
	\definecolor{YELLOW}{cmyk}{0,0,1,0}
}

\@ifundefined{definecolor}
{\usepackage{color}}{}
\@ifundefined{definecolor} 
{\usepackage{color}}{}
\makeatother
\usepackage{babel}

\allowdisplaybreaks

\begin{document}
	

	\title{ Parametrization of zero-skewness unpolarized GPDs
	 }
	
	\author{Hossein Vaziri\orcidA{}}
    \email{Hossein.Vaziri@shahroodut.ac.ir
    }
	
	\author {Mohammad Reza Shojaei\orcidB{}}
	\email{Shojaei.ph@gmail.com}

\affiliation {
	Department of Physics, Shahrood University of Technology, P. O. Box 36155-316, Shahrood, Iran }

	\date{\today}

	%
	%

	%
	\begin{abstract}\label{abstract}
		Recent parameterizations of parton distribution functions (PDFs) have led to the determination of the gravitional form factors of the nucleon's dependence on generalized parton distributions of nucleons in the limit $\xi$$\to 0$. This paper aims to obtain the flavor division of nucleon electromagnetic and gravitional form factors using the VS24 Ansatz and two PDFs at $N^3L0$ approximation in GPDs. The PDFs and GPDs formalism enable the calculation of various form factors of nucleons in different approximations, as well as the calculation of the electric radius of nucleons. The study, despite its high approximation complexity, enhances the accuracy of calculations and brings them closer to the experimental values.

	\end{abstract}
	%

	
	\maketitle

	%

	\section{INTRODUCTON}\label{sec:sec1}
	
	The study of quark and gluon-compound particles can supply details about the nature of strong interactions. The spatial and momentum distributions of constituents in a hadron are not fully understood due to theoretical and calculational limitations, primarily relying on measurements~\cite{ZEUS:2020ddd}.
	Deep-inelastic scattering (DIS) and hard proton-proton high-energy collisions involve scattering through the partonic constituents of the hadron. A set of universal parton distribution functions (PDFs) is necessary to predict the rates of various processes. The distributions can be best determined through global fits to all available DIS and hard-scattering data, performed at leading-order (LO), next-to-leading order (NLO), next-to-next-to-leading order ($N^2LO$), or next-to-next-to-next leading order ($N^3LO$) in the strong coupling $\alpha_s$. Over the past few years, there has been a significant enhancement in the precision and kinematic range of experimental measurements for various processes, along with the emergence of new data types. The reliability of global analyses has been enhanced by significant theoretical developments~\cite{Thorne:2007bt}. 
	
Several groups have previously studied the parton distribution functions (PDFs) ~\cite{Masjuan:2014sua,Halzen,Radyushkin:2011dh,Rosenbluth:1950yq,Guidal:2004nd,RezaShojaei:2016oox,Burkardt:2002hr,Miller:1990iz,Guidal:2013rya}. The old form of describing hadron structure, parton distributions, relied on the B-jorken longitudinal variable $x$, while more complex functions, generalized parton distributions (GPDs), depended on $x$, momentum transfer $t$, and skewness parameter $\xi$~\cite{Selyugin:2023hqu, SattaryNikkhoo:2018gzm, HajiHosseiniMojeni:2022okc,Hashamipour:2021kes}. The unique characteristic of GPDs ($x$, $\xi$, $t$) is that the integration of different momenta of GPDs over $x$ gives us different hadron form factors such as electromagnetic and gravitional form factors~\cite{Muller:1994ses,Ji:1996ek,Radyushkin:1997ki}. The $x$ dependence of GPDs is primarily determined by standard parton distribution functions (PDFs), which are derived from deep-inelastic process analysis~\cite{Vaziri:2023xee,HajiHosseiniMojeni:2022tzn,Nikkhoo:2015jzi}.
	
This section uses one Ansatz that illustrates how $x$ and $t$ rely on generalized parton distributions (GPDs)~\cite{Vaziri:2024fud}. We select two parton distribution functions at $N^3LO$ approximation~\cite{Khorramian:2009xz, Blumlein:2006be, Blumlein:2021lmf}, combine them into our Ansatz, and demonstrate a favorable agreement with experimental data for the calculation of form factors and radii of nucleons. Using GPDs, we can calculate the form factors as well as the radius of the nucleons, so the paper's content consists of the following: In Sec.(\ref{sec:sec2}), we introduce the PDFs. This chapter provides a comprehensive explanation of the general method and form of obtaining these functions. The study includes GPDs and hadron form factors of nucleons, along with related remarks, as presented in Sec.(\ref{sec:sec3}). This section provides an explanation of the formalism of GPDS and the calculation method of form factors. The gravitational form factors of the nucleons based on various Ansatsez and JHA21 PDFs at $N^3LO$ approximation~\cite{Blumlein:2006be, Blumlein:2021lmf} are presented in Sec.(\ref{sec:sec4}) . Sec.(\ref{sec:sec5}) presents the electric radii of nucleons using the combination of the VS24 ansatz~\cite{Vaziri:2024fud} and JHA21 PDFs~\cite{ Blumlein:2006be, Blumlein:2021lmf}. The results and conclusions of our study are presented in Sec.(\ref{sec:conclusion}).

      \section{PARTON DISTRIBUTION FUNCTIONS
      }\label{sec:sec2}

\begin{figure*}
	\includegraphics[clip,width=0.45\textwidth]{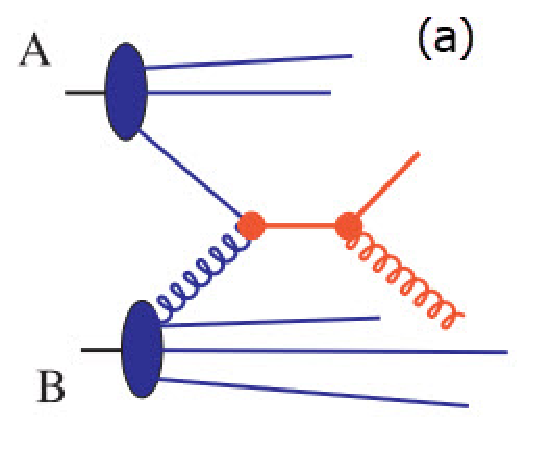}
	\includegraphics[clip,width=0.45\textwidth]{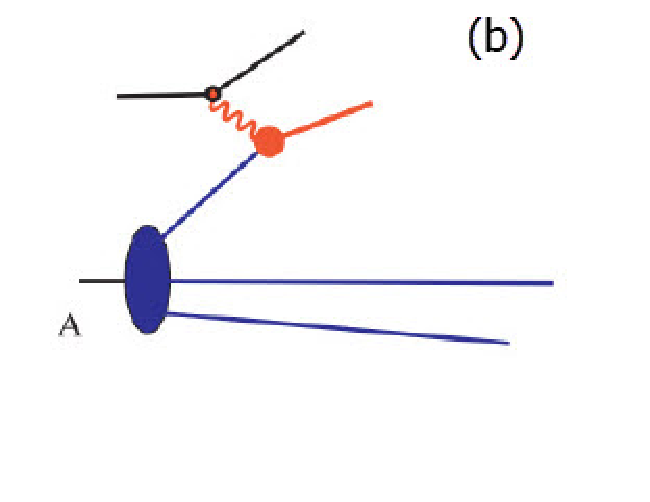}\
	\caption{\footnotesize $a)Hadron A+ Hadron  B \rightarrow 2 Partons$~\cite{Soper:1996sn}. \,\,\,\,\,\,\,\,\,\,\,\,\,\,\,\,\,\,\,\,\,\,\,\,\,\
		b)  Deeply inelastic scattering~\cite{Soper:1996sn}.}
	\label{fig:1}
\end{figure*}

  Parton distribution functions (PDFs) are a necessary ingredient in the calculation of particle cross-sections at collider experiments with hadron beams. The explanation of hard processes with one or two hadrons in the initial state requires an understanding of parton distribution functions. 
  Knowledge of parton distribution functions is necessary for the description of hard processes
  with one or two hadrons in the initial state. (Fig.~\ref{fig:1}). 
  
The cross section of a hadron in its initial state, as seen in deeply inelastic lepton scattering at HERA (Fig.~\ref{fig:1}), consists of the following form ~\cite{Soper:1996sn}:
\begin{equation}
	d\sigma \sim\sum_{a,b}\int dx_A{f}_{a/A}(x_A,\mu) d \widehat {\sigma}.
	\label{eq:sigma2} 
\end{equation}

  There exists a relationship between matrix elements of some local operators and the moments of the parton distribution functions, which are present in the operator product expansion for deeply inelastic scattering. This relation could also be used as the definition~\cite{SattaryNikkhoo:2018odd,Diehl:2003ny}. The technical definition of parton distribution functions is now ready to be studied. There are, in fact, two definitions in current use: the $\overline{MS}$ definition, which is the most commonly used. There is also the DIS definition in which deeply inelastic scattering plays a privileged role~\cite{Soper:1996sn}. The resulting PDFs depend on the selected input data, the order in which the perturbative QCD computation is carried out, the assumptions regarding the PDFs, the handling of heavy quarks, and the treatment of the uncertainties. Presently, the determination of PDFs is carried out by several groups, namely MSTW \cite{Martin:2009iq}
, CTEQ \cite{Nadolsky:2008zw}, NNPDF \cite{Ball:2008by}, HERAPDF \cite{H1:2009pze}, AB(K)M \cite{Alekhin:2009ni}, and GJR \cite{MadrizAguilar:2007dzo}. According to the mentioned methods, PDFs are obtained, which is a function of the average momentum fraction $x$.  Usually, the form of these functions for the $u$ and $d$ quarks is as follows: 

\begin{equation}
	xu_v=A_ux^{\eta_1}(1-x)^{\eta_2}(1+\epsilon_u x^{\eta_3}+\gamma_u x),
	\label{eq:uv}
\end{equation}

\begin{equation}
	xd_v=A_dx^{\eta_4}(1-x)^{\eta_5}(1+\epsilon_d x^{\eta_6}+\gamma_d x),
	\label{eq:dv}
\end{equation}
where the coefficients are obtained by fitting with the experimental data. 
For example, the JHA21 parton distribution functions at $N^3LO$ approximation are as follows ~\cite{ Blumlein:2006be, Blumlein:2021lmf}: 
\begin{equation}
	xu_v=0.261x^{0.298}(1-x)^{4.032}(1+6.042x^{0.5}+35.492x),
	\label{eq:KUV1}
\end{equation}

\begin{equation}
	xd_v=1.085x^{0.5}(1-x)^{5.921}(1-3.618x^{0.5}+16.414x).
	\label{eq:KdV1}
\end{equation}

And the KKA10 parton distribution functions at $N^3LO$ approximation are~\cite {Khorramian:2009xz}:
\begin{equation}
	xu_v=3.41356 x^{0.298}(1-x)^{3.76847}(1+0.1399x^{0.5}-1.12x),
	\label{eq:6}
\end{equation}

\begin{equation}
	xd_v=5.10129x^{0.79167}(1-x)^{4.02637}(1+0.09x^{0.5}+1.11x).
	\label{eq:7}
\end{equation}
 
\begin{figure*}[h!]
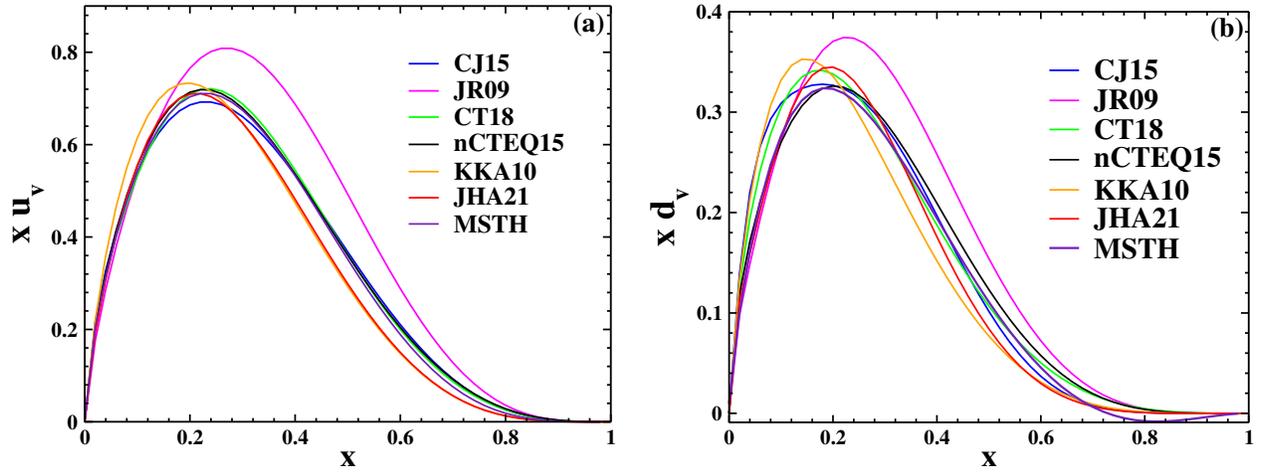

	\includegraphics[clip,width=0.45\textwidth]{xuv.eps}
	\hspace*{2mm}
	\includegraphics[clip,width=0.45\textwidth]{xdv.eps}
	\vspace*{3mm}\\
	
	\caption{\footnotesize  The $xu_v$ and $xd_v$ of the CJ15 at NLO approximation ~\cite{Accardi:2016qay}, JR09~\cite{Jimenez-Delgado:2008orh} at NNLO approximation, CT18~\cite{Hou:2019efy} at NNLO approximation, nCTEQ15~\cite {Kovarik:2015cma}, KKA10~\cite {Khorramian:2009xz}, JHA21~\cite{Blumlein:2006be,Blumlein:2021lmf}, and MSTH~\cite{McGowan:2023mf} PDFs at NNNLO approximation as a function of $x$. } 
	\label{fig:uvdv}
\end{figure*}

 Fig. \ref{fig:uvdv} shows some examples of different PDF charts in different approximations.

  \section{GPDs AND HADRON FORM FACTORS
  }\label{sec:sec3}
 
  The hadron form factors are linked to the GPDs ($x$, $~\xi$, $t$) using the sum rules~\cite{Guidal:2004nd, Selyugin:2009ic}:
 
  \begin{equation}
  	F_{1}(t)=\sum_{q} e_{q}\int_{-1}^{1}dx {H}^{q}(x,t,\xi),
  	\label{eq:f0}
  \end{equation}

  \begin{equation}
  	F_{2}(t)=\sum_{q} e_{q}\int_{-1}^{1}dx {E }^{q}(x,t,\xi),
  	\label{eq:f2}
  \end{equation}
When the momentum is transverse and located in the space-like region, the value of $\xi$ is equal to zero. In the range of $0 < x < 1$, the integration region can be reduced. By revising the elastic form factors, we can obtain:

 \begin{equation}
	F_{1}(t)=\sum_{q} e_{q}\int_{0}^{1}dx \mathcal{H}^{q}(x,t,\xi=0),
	\label{eq:f1}
\end{equation}

\begin{equation}
	F_{2}(t)=\sum_{q} e_{q}\int_{0}^{1}dx \mathcal{E }^{q}(x,t,\xi=0),
	\label{eq:f22}
\end{equation}

In the limit t $\rightarrow$ 0, the functions $H^q(x, t)$ decrease to usual quark densities in the proton:\\
  \begin{equation}
\mathcal{H}^u(x,t=0)= u_v(x) ,  \,\,\,\,\,\,\,\,\,\,\    \mathcal{H}^d(x,t=0)= d_v(x),
	\label{eq:1}
  \end{equation}

with the integrals\\

  \begin{equation}
\int_{0}^{1}  u_v(x) dx=2 ,\,\,\,\,\,\,\,\,\,\,\    \int _{0}^{1} d_v(x) dx=1 .
	\label{eq:2}
  \end{equation}
normalized to the proton's $u$ and $d$ valence quark numbers.
These factors result in the determination of certain parameters. For instance, the values $\kappa_u = 1.673$ and $\kappa_d = -2.033$ are obtainable. Moreover, the normalization integral for the mathematical constant $\int_{0}^{1}\mathcal{H}_{q}(x,0)$ takes on particular values for the nucleons: $F^P_1(0) = 1$ for the proton and $F_1^n(0) = 0$ for the neutron. 

The functions $\mathcal{H}(x)$ and $\varepsilon(x)$ differ in each of the models that have been suggested. To yield a faster reduction with $t$, the $x \to 0$ limit of $\varepsilon(x)$ should contain additional powers of $(1-x)$, compared with $\mathcal{H}(x)$~\cite{Guidal:2004nd,Selyugin:2009ic,Mojeni:2020rev}. Hence, we have:
\begin{eqnarray}
	\varepsilon _{u}(x) &=&\frac{\kappa _{u}}{N_{u}}(1-x)^{\eta _{u}}u_{v}(x), \nonumber\\
	\varepsilon _{d}(x) &=&\frac{\kappa _{d}}{N_{d}} (1-x)^{\eta _{d}}d_{v}(x),\label{eq:Eud1}
\end{eqnarray}

where the normalization factors $N_u$ and $N_d$ are determined as~\cite{Guidal:2004nd}:

\begin{eqnarray}
	N_{u} &=&\int_{0}^{1}dx(1-x)^{\eta _{u}}u_{v}(x), \\
	N_{d} &=&\int_{0}^{1}dx(1-x)^{\eta _{d}}d_{v}(x).  \nonumber
	\label{eq:Nud}
\end{eqnarray}

The proton value must be $F_2^p(0)=\kappa_p = 1.793$, and the neutron value must be $F_2^n(0)=\kappa_n = -1.913$ to satisfy the limits on $\kappa_{q}$.

\begin{equation}
	\kappa _{q}=\int_{0}^{1}dx\varepsilon _{q}(x).
\end{equation}

\begin{eqnarray}
	\kappa_u=\kappa_n+2\kappa_p,\nonumber\\
	\kappa_d=2\kappa_n+\kappa_p.
\end{eqnarray}

One of the most important techniques for examining the structure of nucleons is the use of generalized parton distributions, or GPDs~\cite{Diehl:2003ny,Shojaei:2015oia,SattaryNikkhoo:2018odd}.

In order to satisfy the requirements outlined in Eq. (\ref{eq:Eud1}), the nucleon form factor data was fitted to obtain the values of $\eta_u$ and $\eta_d$. We must adjust our models in order to increase the agreement with the data for large $-t$.

The extended ER Ansatz~\cite{Guidal:2004nd}, modified Gaussian (MG) Ansatz~\cite{Selyugin:2009ic}, HS22 Ansatz~\cite{HajiHosseiniMojeni:2022tzn}, M-HS22 Ansatz~\cite{Vaziri:2023xee}, and VS24 Ansatz~\cite{Vaziri:2024fud} are among the Ansatzes with a $t$ dependency that will be presented in this paper.
In this section, first we introduce different Ansatzes. The ER Ansatz is~\cite{Guidal:2004nd}: 

\begin{equation}
	\mathcal{H}^{q}(x,t)=q_{v}(x)x^{-\alpha ^{\prime }(1-x)t},
	\label{eq:HER}
\end{equation}
\begin{equation}
	\varepsilon _{q}(x,t)=\varepsilon _{q}(x)x^{-\alpha ^{\prime }(1-x)t}.
	\label{eq:EER}
\end{equation}

And the modified Gaussian Ansatz (MG) is followed as~\cite{Selyugin:2009ic}:
\begin{equation}
	\mathcal{H}^{q}(x,t)=q_{v}(x)\exp \left[ \alpha \frac{(1-x)^{2}}{x^{m}}t%
	\right] ,
	\label{eq:HMG}
\end{equation}

\begin{equation}
	\varepsilon _{q}(x,t)=\varepsilon _{q}(x)\exp \left[ \alpha \frac{(1-x)^{2}}{x^{m}}t%
	\right] .
	\label{eq:EMG}
\end{equation}

\begin{figure*}
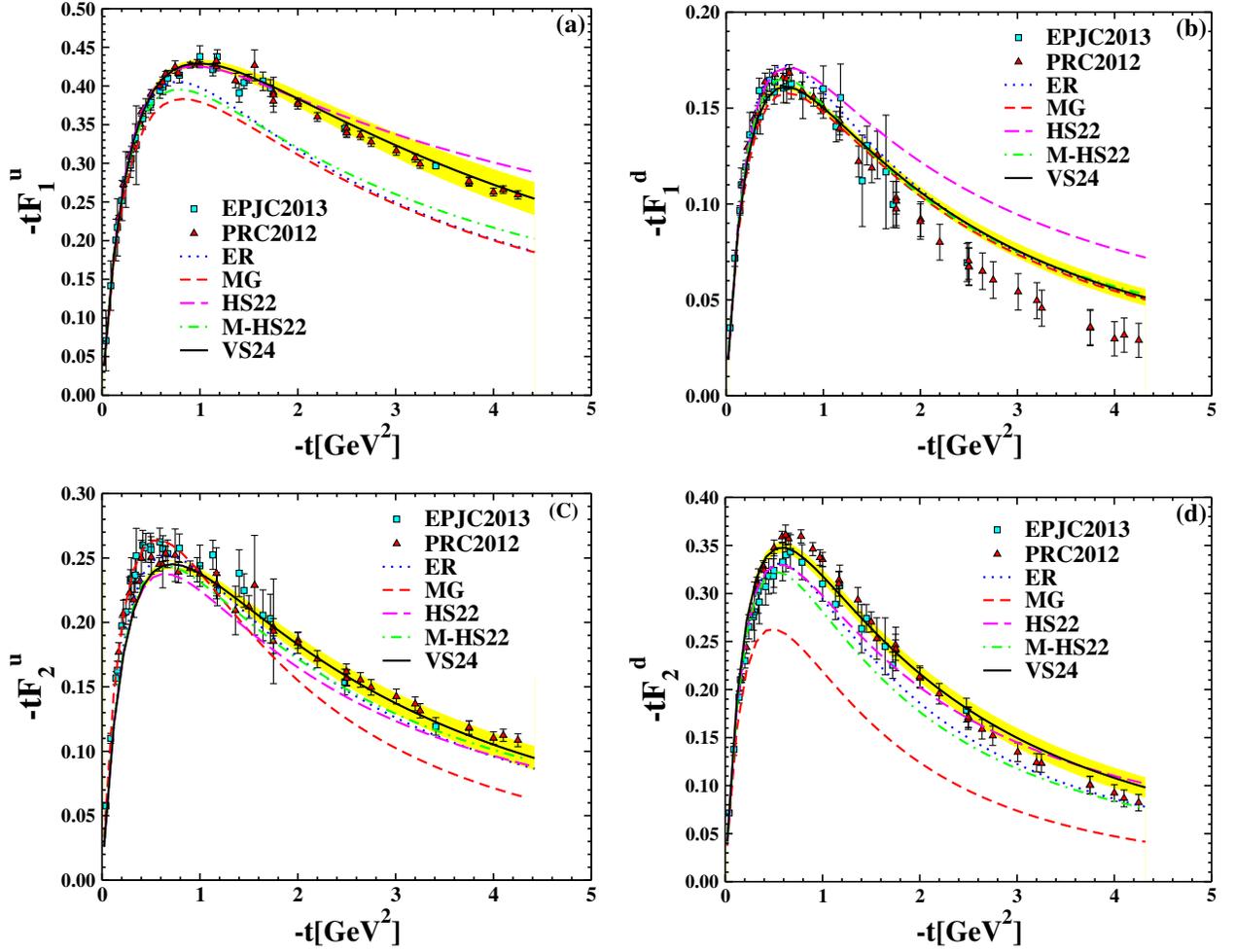

	\includegraphics[clip,width=0.45\textwidth]{tf1u.eps}
	\hspace*{2mm}
	\includegraphics[clip,width=0.45\textwidth]{tf1d.eps}
	\vspace*{3mm}\\
	\includegraphics[clip,width=0.45\textwidth]{tf2u.eps}
	\hspace*{2mm}
	\includegraphics[clip,width=0.45\textwidth]{tf2d.eps}
	\vspace*{1.5mm}
	
	\caption{\footnotesize  The $F_1^{u,d}$ and $F_2^{u,d}$ are multiplied by $t$ as a function of $-t$. Comparison of the ER Ansatz~\cite{Guidal:2004nd}, the MG Ansatz\cite{Selyugin:2009ic}, the HS22 Ansatz~\cite{HajiHosseiniMojeni:2022tzn}, the M-HS22 Ansatz~\cite{Vaziri:2023xee} with the VS24 Ansatz~\cite{Vaziri:2024fud}. All of them make use of the KKA10 PDF~\cite{Khorramian:2009xz}. Experimental data from~\cite{Qattan:2012zf} (triangle up),~\cite{Cates:2011pz} (circle), and ~\cite{Diehl:2013xca} (square) served as a basis for the extracted points.}
	\label{fig:tfud1}
\end{figure*}

\begin{figure*}
	\includegraphics[clip,width=0.45\textwidth]{tf1u1.eps}
	\hspace*{2mm}
	\includegraphics[clip,width=0.45\textwidth]{tf1d1.eps}
	\vspace*{3mm}\\
	\includegraphics[clip,width=0.45\textwidth]{tf2u1.eps}
	\hspace*{2mm}
	\includegraphics[clip,width=0.45\textwidth]{tf2d1.eps}
	\vspace*{1.5mm}
	\caption{\footnotesize  The $F_1^{u,d}$ and $F_2^{u,d}$ are multiplied by $t$ as a function of $-t$. Comparison of the ER Ansatz~\cite{Guidal:2004nd}, the MG Ansatz\cite{Selyugin:2009ic}, the HS22 Ansatz~\cite{HajiHosseiniMojeni:2022tzn}, the M-HS22 Ansatz~\cite{Vaziri:2023xee} with the VS24 Ansatz~\cite{Vaziri:2024fud}. The JHA21 PDF~\cite{ Blumlein:2006be, Blumlein:2021lmf} is used in all of them. The extracted points are based on experimental data from ~\cite{Qattan:2012zf} (triangle up),~\cite{Cates:2011pz} (circle), and ~\cite{Diehl:2013xca} (square).}
	\label{fig:tfuBL}
\end{figure*}

\begin{figure*}
	\includegraphics[clip,width=0.45\textwidth]{tf1p1.eps}
	\hspace*{2mm}
	\includegraphics[clip,width=0.45\textwidth]{tf1n1.eps}
	\vspace*{3mm}\\
	\includegraphics[clip,width=0.45\textwidth]{tf2p1.eps}
	\hspace*{3mm}
	\includegraphics[clip,width=0.45\textwidth]{tf2n1.eps}
	\vspace*{1.5mm}
	\caption{\footnotesize  The $F_1^{p,n}$ and $F_2^{p,n}$ are multiplied by $t$ as a function of $-t$. Comparison of the ER Ansatz~\cite{Guidal:2004nd}, the MG Ansatz~\cite{Selyugin:2009ic}, the HS22 Ansatz~\cite{HajiHosseiniMojeni:2022tzn}, the M-HS22 Ansatz~\cite{Vaziri:2023xee}, and the VS24 Ansatz~\cite{Vaziri:2024fud}. The JHA21 PDF~\cite{Blumlein:2006be, Blumlein:2021lmf} is used in all of them. The extracted points are based on experimental data from ~\cite{Qattan:2012zf} (triangle up),~\cite{Cates:2011pz} (circle), and ~\cite{Diehl:2013xca} (square).}
	\label{fig:tfpBL}
\end{figure*}

\begin{figure*}
	\includegraphics[clip,width=0.45\textwidth]{GEp1.eps}
	\hspace*{2mm}
	\includegraphics[clip,width=0.45\textwidth]{GEn1.eps}
	\vspace*{3mm}\\
	\includegraphics[clip,width=0.45\textwidth]{GMP1.eps}
	\hspace*{2mm}
	\includegraphics[clip,width=0.45\textwidth]{GMn1.eps}
	\vspace*{1.5mm}
	\caption{\footnotesize    The $G_E^{p,n}$ and $G_M^{p,n}$ as a function of $-t$. Comparison of the ER Ansatz~\cite{Guidal:2004nd}, the MG Ansatz~\cite{Selyugin:2009ic}, the HS22 Ansatz~\cite{HajiHosseiniMojeni:2022tzn}, the M-HS22 Ansatz~\cite{Vaziri:2023xee}, and the VS24 Ansatz~\cite{Vaziri:2024fud}. The JHA21 PDF~\cite{ Blumlein:2006be, Blumlein:2021lmf} is used in all of them. The extracted points are based on experimental data from ~\cite{Qattan:2012zf} (triangle up).}
	
	\label{fig:GEM BL}
\end{figure*}

The free parameters for the modified Gaussian (MG) are $m=0.45$ and $\alpha=1.15$, and for the extended ER, $\alpha^{\prime}=1.09$.

We introduced the HS22 Ansatz and the M-HS22 Ansatz in Refs.~\cite{HajiHosseiniMojeni:2022tzn} and ~\cite{Vaziri:2023xee}, respectively, which have more parameters than the ER and MG Ansatzes. 

We vary our previous M-HS22 Ansatz~\cite{Vaziri:2023xee} for the GPDs by adding two new parameters, $m^{\prime\prime}$ and $\gamma$, and introduce the VS24 Ansatz~\cite{Vaziri:2024fud}, which is as follows:

\begin{equation}
	\mathcal{H}_{q}(x,t)=q_{v}\;\exp [-\alpha^{\prime \prime \prime} t(1-x)^\gamma\ln (x)+\beta x^{m^\prime}\ln (1-bt)],
	\label{eq:H}
\end{equation}
\begin{equation}
	\varepsilon _{q}(x,t)=\varepsilon _{q}(x)\;\exp [-\alpha^{\prime \prime \prime} t (1-x)^\gamma\ln (x)+\beta
	x^{m^\prime}\ln (1-bt)],
	\label{eq:E}
\end{equation}

Using the VS24 ansatz in combination with KKA10~\cite{Khorramian:2009xz} and JHA21~\cite{Khorramian:2009xz, Blumlein:2006be, Blumlein:2021lmf} PDFs, we calculate the form factors of the $u$ and $d$ quarks based on the formalism described earlier. 

The Dirac and Pauli form factors of the $u$ and $d$ quarks that were obtained by KKA10 \cite{Khorramian:2009xz} and JHA21 PDFs~\cite{Blumlein:2006be, Blumlein:2021lmf} are shown in Fig.~\ref{fig:tfud1} and Fig.~\ref{fig:tfuBL}, respectively, as functions of $-t$. By combining five different types of Ansatz with two PDFs, the form factors of the $u$ and $d$ quarks are calculated.

As is evident from these figures, the form factors obtained from the VS24 Ansatz~\cite{Vaziri:2024fud} combined with JHA21 PDFs~\cite{Blumlein:2006be,Blumlein:2021lmf} show better agreement with the form factors obtained from electron-proton inelastic scattering experiments~\cite{Qattan:2012zf,Cates:2011pz,Diehl:2013xca} than those obtained with KKA10 PDFs~\cite{Khorramian:2009xz}.

The proton and neutron Dirac form factors are defined as:
\begin{equation}
	F_{1}^{p}(t)= e_u F_{1}^{u}(t)+e_d F_{1}^{d}(t)  ,
\end{equation}

\begin{equation} 
	F_{1}^{n}(t)= e_d F_{2}^{u}(t)+e_d F_{2}^{d}(t).
	\label{eq:p}
\end{equation}

And therefore the Pauli form factors are:
\begin{equation}
	F_{2}^{p}(t)= e_u F_{2}^{u}(t)+e_d F_{2}^{d}(t)  ,
\end{equation}

\begin{equation} 
	F_{2}^{n}(t)= e_d F_{2}^{u}(t)+e_u F_{2}^{d}(t).
	\label{eq:n}
\end{equation}

where $e_u = 2/3$ and $e_d = -1/3$ are the corresponding quark electric charges. As a result, the t-dependence of the GPDs ($x$, $\xi = 0$, $t$) can be determined from the analysis of the nucleon form factors for which experimental data exist in a wide region of momentum transfer.

The diagrams of form factors of nucleons as a function of $-t$ are drawn in Fig.~\ref{fig:tfpBL}, and their calculations are obtained from the combination of the VS24 Ansatz~\cite{Vaziri:2024fud} and the different PDFs.

\begin{table}[H]
	\begin{center}
		\caption{{\footnotesize Coefficients for calculations of the Eqs.(\ref{eq:H}) and (\ref{eq:E}). We have used the VS24 Ansatz~\cite{Vaziri:2024fud}. The parameters of $b$ and $m^{\prime}$ are fixed, and the others have been calculated by fitting.}
			\label{tab:tabpdf}}
		\vspace*{0.4mm}
		\begin{tabular}{cccccc}
			\hline\hline\
			\\& {\hspace{1mm}}   & {\hspace{1mm}} \textbf{KKA10 PDFs~\cite{Khorramian:2009xz}}& {\hspace{1mm}}\textbf{JHA21 PDFs~\cite{ Blumlein:2006be, Blumlein:2021lmf}}&     \\   \hline
			\\    & $\alpha ^{\prime\prime}$$$ & {\hspace{2mm}}$1.3742\pm 7.16734\times 10^{-3}$& {\hspace{1mm}}$1.3473\pm 7.18952\times 10^{-3}$&   \\\\
			$   $  & {\hspace{1mm}}$\beta$ & {\hspace{1mm}}$1.52378\pm 2.8392\times 10^{-2}$& {\hspace{1mm}}$1.51418\pm 2.5564\times 10^{-2}$&     \\\\ 
			& {\hspace{1mm}}$\gamma$ & {\hspace{1mm}}$0.0570108\pm 1.49425\times 10^{-2}$& {\hspace{1mm}}$2.9823\pm 1.2596\times 10^{-2}$&{\hspace{1mm}}&  \\\\
			$$  & {\hspace{1mm}}$\eta_u$ & {\hspace{1mm}}$0.71207\pm 9.909\times 10^{-3}$& {\hspace{1mm}}$0.6931\pm 9.8663\times 10^{-3}$& {\hspace{1mm}}& \\\\
			& {\hspace{1mm}}$\eta_d$ & {\hspace{1mm}}$0.19248\pm 1.606\times 10^{-2}$& {\hspace{1mm}}$0.2782\pm 1.7735\times 10^{-2}$& {\hspace{1mm}}&    \\\\ 
			$$  & {\hspace{1mm}}$b$ & {\hspace{1mm}}$2$&{\hspace{1mm}}$2$    \\\\ 
			$$  & {\hspace{1mm}}$m^{\prime}$ & {\hspace{1mm}}$0.65$&  {\hspace{1mm}}$0.65$  \\\\

			\hline\hline						
		\end{tabular}
	\end{center}
\end{table}

The Sachs form factors can be derived from $F_{1}(t)$ and $F_{2}(t)$ in the manner that follows ~\cite{Ernst:1960zza,Nikkhoo:2017won,Guidal:2004nd}:

\begin{equation}
	G_{E}^{N}(t)= -\tau F_{2}(t)+F_{1}(t)    ,\;\;\;G_{M}^{N}(t)=F_{2}(t)+F_{1}(t).
	\label{eq:GN}
\end{equation}
where $t=Q^2$ is the four-momentum transfer of the virtual photon and $\tau \equiv -t / 4 M_N^2$.
Sach suggests that the form factors $G_{E}$ and $G_{M}$ may have more fundamental significance compared to the $F_1$ and $F_2$ in interpreting the spatial distributions of charge and magnetization inside the nucleon, as they are related to the four-momentum transfer of the virtual photon~\cite{Ernst:1960zza}.

Since the errors on $F_1$ and $F_2$ are typically more correlated and larger compared to the errors on $G_E$ and $G_M$, however, we studied the $G_E$ and $G_M$ for nucleons. There are differences between the down- and up-quark distributions~\cite{Drell:1969km}:
\begin{equation}
	G^p_{E,M}(Q^2)=\frac{2}{3}G^u_{E,M}(Q^2)-\frac{1}{3}G^d_{E,M}(Q^2),
\end{equation}

\begin{equation} 
	G^n_{E,M}(Q^2)=\frac{2}{3}G^d_{E,M}(Q^2)-\frac{1}{3}G^u_{E,M}(Q^2).
\end{equation}

The form factors in the combination of the VS24 Ansatz~\cite{Vaziri:2024fud} and the JHA21 PDFs~\cite{Blumlein:2006be, Blumlein:2021lmf} were compared to those obtained from scattering experiments in the Rosenbluth equation at $q^2=0$, as shown in Tab. (\ref{tab:t=0}).

\begin{table}[h]
	\begin{center}
		\caption{{\footnotesize The values of $G^p_{E,M}$ and $G^n_{E,M}$ at $t=0$. }
			\label{tab:t=0}}
		\vspace*{1.5mm}
		\begin{tabular}{ccc}
			\hline\hline\
			\\	$G^{P,n}_{E,M}(0)$ & {\hspace{4mm}}EXP.DATA~\cite{Hohler:1976ax} & {\hspace{4mm}}JHA21~\cite{ Blumlein:2006be, Blumlein:2021lmf}+VS24~\cite{Vaziri:2024fud}    \\   \hline
			\\$G^P_{E}(0)$   &{\hspace{10mm}} 1 & {\hspace{10mm}} 1 \\
			\\
			$G^P_{M}(0)$   &{\hspace{10mm}}   +2.79  &{\hspace{10mm}} +2.65867 \\   
			\\
			$G^n_{E}(0)$   &{\hspace{10mm}} 0 &{\hspace{10mm}}$3.2\times 10^{-13}$  \\ 
			\\
			
			$G^n_{M}(0)$ &{\hspace{10mm}}   -1.91 &{\hspace{10mm}}-1.94332 \\
			\hline\hline						
		\end{tabular}
	\end{center}
\end{table}

Studying the contributions of up and down quarks to the nucleon's form factors can provide insights into its fundamental structure and dynamics~\cite{Beck:2001yx}. Fig.~(\ref{fig:GEM BL}) shows the electric and magnetic form factors of the nucleons, derived from JHA21 PDFs~\cite{Blumlein:2006be, Blumlein:2021lmf}, and plotted as functions of $-t$. 

This graph shows that the form factors obtained from the combination of the VS24 Ansatz~\cite{Vaziri:2024fud} and the JHA21 PDFs~\cite{Blumlein:2006be, Blumlein:2021lmf} are more consistent with experimental data~\cite{Qattan:2012zf, Cates:2011pz, Diehl:2013xca} than other combinations of the Ansatzes with JHA21 PDFs~\cite{Blumlein:2006be, Blumlein:2021lmf}. It can be seen that especially for $G_E^n$, it is even better than the form factors calculated by the combination of the VS24 Ansatz and KKA10 PDFs in Ref.~\cite{Vaziri:2024fud}.

	\section{ Gravitational Form Factors of Nucleons (Quark and Gluon Contributions) }\label{sec:sec4}
	There are four GFFs of the proton: $A(q^2)$, $B(q^2)$, $C(q^2)$, and $\bar{C}(q^2)$. The GFFs $A(q^2)$ and $B(q^2)$ are related to the mass and angular momentum distributions of the proton.
	Conservation of the energy-momentum tensor constrains the GFFs $A(q^2)$ and $B(q^2)$, and $\bar{C}(q^2)$; however , $C(q^2)$, also known as the D-term, is not related to any Poincare generator and is unconstrained by such conservation laws. The D-term contributes to the DVCS process when the skewness $\xi$ is nonzero, or when there is a
	longitudinal momentum transfer from the initial state proton to the final state proton ~\cite{More:2021stk}.
	Taking the matrix elements of the energy-momentum tensor $T_{\mu \nu}$ instead of the electromagnetic current $J^{\mu}$, one can obtain the gravitational form factors of quarks, which are related to the second rather than the first moments of GPDs ~\cite{GarciaMartin-Caro:2023klo}:
	\begin{eqnarray}
		A_{q}(t)=\int_{-1}^{1}dx \ x \ {H}^{q}(x,t,\xi), \\
				\label{eq:33}
		B_{q}(t)=\int_{-1}^{1}dx \ x \ {E}^{q}(x,t,\xi).  
				\label{eq:34}
	\end{eqnarray}

For $\xi=0$, the valence contribution to the gravitational form factor is as follows:
\begin{eqnarray}
	A_{q}(t)=\int_{0}^{1}dx x {H}^{q}(x,t), \\
	B_{q}(t)=\int_{0}^{1}dx x {E}^{q}(x,t).  
	\label{eq:36}
\end{eqnarray}
This representation, combined with our model (we use here the first variant of parameters describing the experimental data obtained by the polarization method), allows to calculate the gravitational form factors of valence quarks and their contribution (being just their sum) to gravitational form factors of nucleons.
The $t$ dependency of $A_{u+d}(t)$ is the same for our GPDs with varying PDFs. These contributions add up to $A_{u+d}(t) \approx 0.45$ at $t = 0$.

\begin{figure*}[h]
		\includegraphics[clip,width=0.45\textwidth]{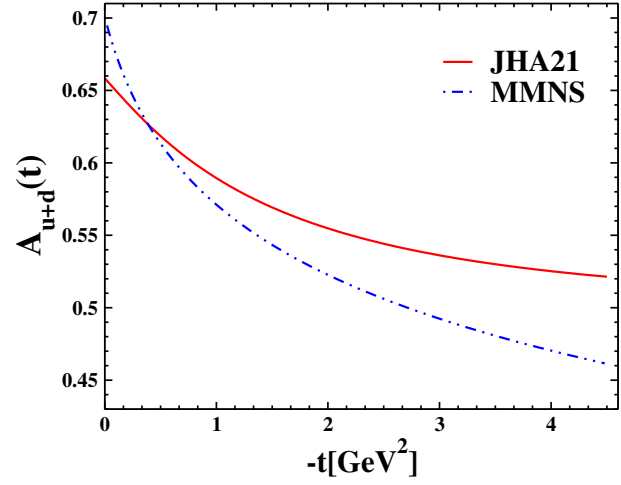}
		\hspace*{2mm}
		\caption{\footnotesize    The proton GFF $A_{u+d}$ as a function of $-t$. The JHA21 PDF in the NNNLO approximation~\cite{Blumlein:2006be, Blumlein:2021lmf}, with combination of the VS24 Ansatz~\cite{Vaziri:2024fud}. The results of MMNS (Lattice QCD)~\cite{More:2021stk} are plotted.}
		\label{fig:Au+d}
		
	\end{figure*}

\begin{figure*}[h]
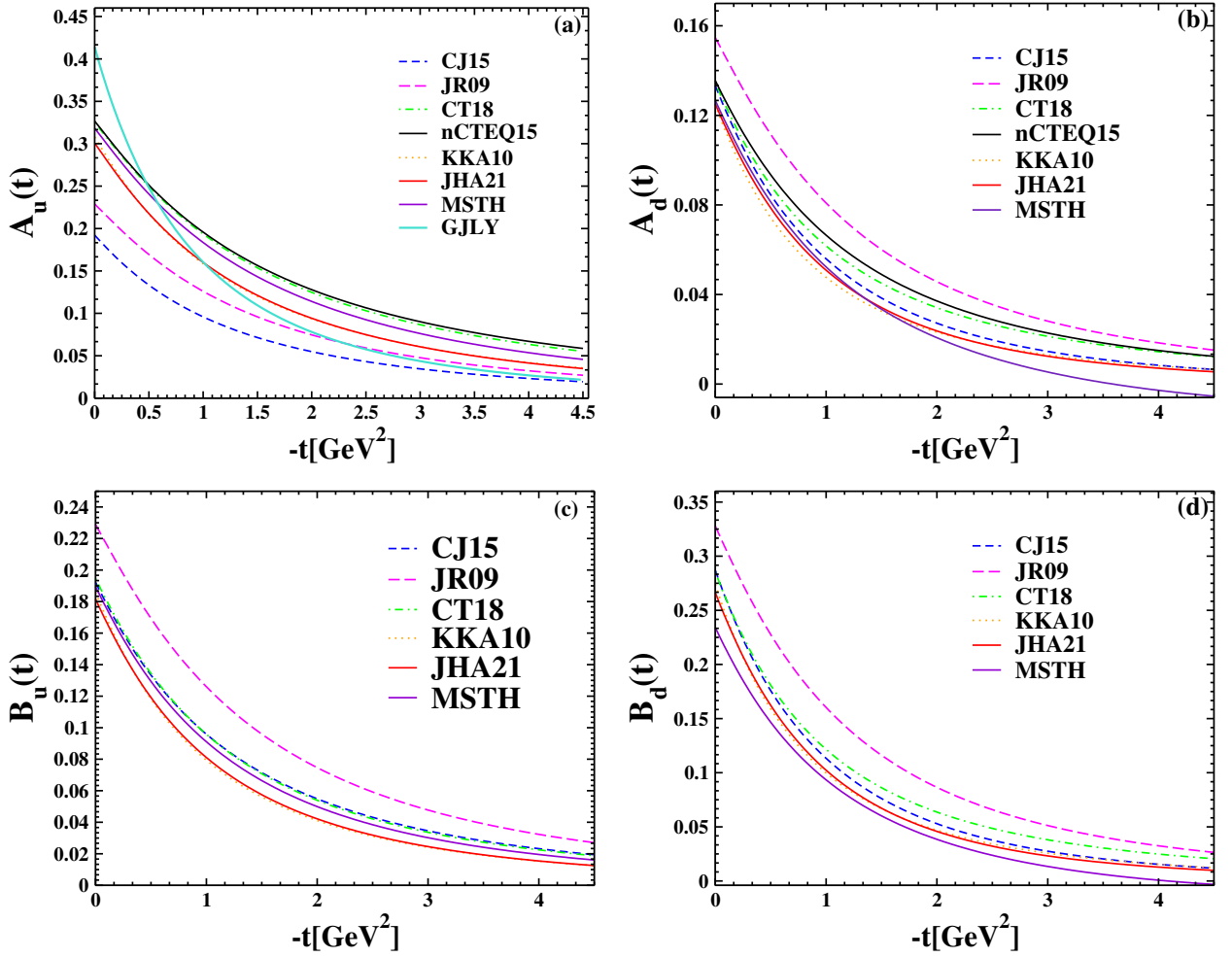

		\includegraphics[clip,width=0.45\textwidth]{Au.eps}
		\hspace*{2mm}
		\includegraphics[clip,width=0.45\textwidth]{Ad.eps}
		\vspace*{3mm}\\
		\includegraphics[clip,width=0.45\textwidth]{Bu.eps}
		\hspace*{2mm}
		\includegraphics[clip,width=0.45\textwidth]{Bd.eps}
		\vspace*{1.5mm}
		\caption{\footnotesize    The $A^{u,d}$ and $B^{u,d}$ as a function of $-t$. The CJ15 parton distribution functions in the NLO approximation~\cite{Accardi:2016qay}, the JR09 parton distribution functions in the NNLO approximation~\cite{Jimenez-Delgado:2008orh}, the CT18 parton distribution functions in the NNLO approximation~\cite{Hou:2019efy}, the nCTEQ15 PDF in the NNLO approximation~\cite{Kovarik:2015cma}, the KKA10 PDF in the NNNLO approximation~\cite{Khorramian:2009xz}, the JHA21 PDF in the NNNLO approximation~\cite{Blumlein:2006be,Blumlein:2021lmf}, and the MSTH PDF in the NNNLO approximation~\cite{McGowan:2023mf} with combination of the VS24 Ansatz~\cite{Vaziri:2024fud}. The GFF of GJLY~\cite{Guo:2023pqw} is plotted.}
		\label{fig:AqBq}
	
\end{figure*}

These various PDFs for different PDFs are given in Tab.\ref{tab:t=1}. These PDFs are:          

The CJ15 parton distribution functions in the NLO approximation~\cite{Accardi:2016qay}, the JR09 parton distribution functions in the NNLO approximation~\cite{Jimenez-Delgado:2008orh}, the CT18 parton distribution functions in the NNLO approximation~\cite{Hou:2019efy}, the nCTEQ15 PDF in the NNLO approximation~\cite{Kovarik:2015cma}, the KKA10 PDF in the NNNLO approximation~\cite{Khorramian:2009xz}, the JHA21 PDF in the NNNLO approximation~\cite{Blumlein:2006be, Blumlein:2021lmf}, and the MSTH PDF in the NNNLO approximation~\cite{McGowan:2023mf}. Combining these PDFs with VS24 Ansatz~\cite{Vaziri:2024fud}, we calculated the gravitational form factors for $u$ and $d$ quarks and plotted them as a function of $-t$. In Fig.(\ref{fig:AqBq}), we also calculated the $A^{u}$ for the group of GJLY~\cite{Guo:2023pqw} and plotted the form factor of them. Fig. (\ref{fig:Au+d}) shows a comparison of the proton GFF $A_{u+d}$ as a function of $-t$ for MMNS (Lattice QCD)~\cite{More:2021stk} with the JHA21 PDF in the NNNLO approximation~\cite{Blumlein:2006be, Blumlein:2021lmf}, combining the VS24 Ansatz~\cite{Vaziri:2024fud}. It can be seen that the behavior of all graphs of gravitional form factors is similar by combining these PDFs with the desired Ansatz.

		\begin{table}[h]
		\begin{center}
			\caption{{\footnotesize Exact values of $A_{u+d}(t)$ for different PDFs at $t=0$. The VS24 Ansatz~\cite{Vaziri:2024fud} is used in all of them.}
				\label{tab:t=1}}
			\vspace*{1.5mm}
			\begin{tabular}{ccc}
				\hline\hline\
				\\	PDFs & {\hspace{10mm}} $A_{u+d}(t)$ &    \\   \hline
				\\ CJ15~\cite{Accardi:2016qay}   &{\hspace{10mm}} 0.455512 & \\
				JR09~\cite{Jimenez-Delgado:2008orh}    &{\hspace{10mm}}   0.455512  & \\   
				CT18~\cite{Hou:2019efy}   &{\hspace{10mm}}  0.459818 & \\
				nCTEQ15~\cite{Kovarik:2015cma}   &{\hspace{10mm}} 0.46219 & \\ 
				KKA10~\cite{Khorramian:2009xz}   &{\hspace{10mm}} 0.43291 & \\ 
				JHA21~\cite{ Blumlein:2006be, Blumlein:2021lmf}   &{\hspace{10mm}} 0.425903 & \\
				MSTH~\cite{McGowan:2023mf} &{\hspace{10mm}}   0.44496 & \\
				\hline\hline						
			\end{tabular}
		\end{center}
	\end{table}

	\section{The ELECTRIC RADII OF THE NUCLEONS BASED ON VS24 ANSATZ}\label{sec:sec5}
	
The particle radii at zero momentum transfer are determined by the slope of form factors, and the squares of the Dirac radius $<r_{D}^2>$ and charge radius $r_{E,p}^2$ are determined using the following method~\cite{Selyugin:2019dav}:
	 	 
\begin{eqnarray}
	<r_{D}^2>&=&-6 \frac{dF^{p,n}_{1}(t)}{dt} \mid_{t=0},\nonumber\\	 
	<r_{E}^2>&=&-6 \frac{dF^{p,n}_{1}(t)}{dt} \mid_{t=0}+\; \frac{3}{2} \frac{{\kappa_{{n},p}}}{m_{n,p}^2}.
	 \label{eq:r1,p}
\end{eqnarray}

In addition, we used the VS24 Ansatz~\cite{Vaziri:2024fud} to compute the nucleon's Dirac mean squared radii.

\begin{eqnarray}
	<r_{D,p}^{2}>=-6\alpha^{\prime \prime\prime}\int_{0}^{1}dx    
	[e_{u}u_{v}(x)+e_{d}d_{v}(x)]   (1-x)^\gamma\nonumber\\ 
	\ln (x)+\beta x^{m^\prime}\ln(1-bt) \mid_{t=0},\;\;\;\;\;\;\;\;\;\;\;\;\;\;\;\;\;\;\;\;\;\;\;\;\;\;\;
	\label{eq:rPMG2}
\end{eqnarray}

\begin{eqnarray}
	<r_{D,n}^{2}>=-6\alpha^{\prime \prime\prime}\int_{0}^{1}dx    
	[e_{u}d_{v}(x)+e_{d}u_{v}(x)]   (1-x)^\gamma\nonumber\\ 
	\ln (x)+\beta x^{m^\prime}\ln(1-bt) \mid_{t=0}.\;\;\;\;\;\;\;\;\;\;\;\;\;\;\;\;\;\;\;\;\;\;\;\;\;\;\;
	\label{eq:rPMG3}
\end{eqnarray}
The proton charge radius is defined as the slope of the electric form factor in the forward limit $(t = 0)$, and therefore, it doesn't depend on $-t$.  
In Tab. (\ref{tab:tabR}), we display the calculated electric radii of nucleons using the JHA21 PDFs~\cite{Blumlein:2006be, Blumlein:2021lmf} combined with the VS24 Ansatz~\cite{Vaziri:2024fud} and compare them with experimental data obtained from \cite{Xiong:2019umf}.
\captionsetup{belowskip=0pt,aboveskip=0pt}

\begin{table}[h]
	\begin{center}
		\caption{{\footnotesize The proton's electric radii were calculated by using the JHA21 parton distribution functions (PDFs)~\cite{Blumlein:2006be, Blumlein:2021lmf} and VS24 model~\cite{Vaziri:2024fud}. Ref. \cite{Xiong:2019umf} is the source of the data.}
			\label{tab:tabR}}
		\vspace*{1.5mm}
		\begin{tabular}{ccc}
			\hline\hline\
			\\	Ansatz+PDFs & {\hspace{10mm}}$r_{E,p}$  \\   \hline
			\\Exprimental data\cite{Xiong:2019umf}   &{\hspace{10mm}} 0.831 $fm$ \\\\
			VS24~\cite{Vaziri:2024fud}+JHA21~\cite{ Blumlein:2006be, Blumlein:2021lmf}   &{\hspace{10mm}}   0.853154 $fm$  \\

			 \\
			\hline\hline						
		\end{tabular}
	\end{center}
\end{table}

	\section{ RESULTS AND CONCLUSION}\label{sec:conclusion}
	
The paper examines a collection of parametrizations of the zero-skewness unpolarized quark GPDs, $H$ and $E$, with flavor separation between $u$ and $d$ quarks. The parametrizations are constructed using several PDF extractions and fitted to elastic form factor datasets. Gravitational form factors are then derived. PDFs are functions that only depend on $x$ B-jorken (that indicate the average longitudinal momentum fraction of the partons). These functions are obtained using inelastic scattering data of electrons from protons. But GPDs depend on two other quantities, $t$ and $\xi$, in addition to $x$: $t$ is the momentum transfer, and $\xi$ measures the longitudinal momentum transfer in the hard scattering. The basis of our work is the calculation of the form factors of nucleons, of which quarks are also a part. By calculating the electric, magnetic, and gravitional form factors, we can study the structure of nucleons, such as radius, density, etc. We started with the introduction of PDFs and their formalism in Sec.(\ref{sec:sec2}) and explained the method of obtaining these functions using dispersion relation. The general form of these functions is in the form of Eqs. (\ref{eq:uv}) and (\ref{eq:dv}), and some examples of different PDF charts as a function of $x$ at different approximations are shown in Fig. (\ref{fig:uvdv}). As these figures show, the behavior and physics of these functions are the same.As these figures show, the behavior and physics of these functions are the same. Since GPDs show the space inside the nucleons in three dimensions, we discussed GPDs and form factors in Sec.(\ref{sec:sec3}). In the following, we explained the formalism of GPDs in detail in Eqs. (\ref{eq:f0}-- \ref{eq:Nud}), where $F_1$ and $F_2$ are the Pauli and Dirac form factors, respectively. In fact, ${H}^{q}(x,t,\xi=0)$ and ${E }^{q}(x,t,\xi=0)$ have two parts: The first part is the PDF, and the other is a proposed function extracted by fitting with form factors obtained from scattering experiments, using data from Refs.~\cite{Qattan:2012zf, Cates:2011pz, Diehl:2013xca}. The proposed Ansatz is a function that is usually exponential, and the most important of them are as follows: the ER Ansatz~\cite{Guidal:2004nd}, the MG Ansatz\cite{Selyugin:2009ic}, the HS22 Ansatz~\cite{HajiHosseiniMojeni:2022tzn}, the M-HS22 Ansatz~\cite{Vaziri:2023xee}, and the VS24 Ansatz~\cite{Vaziri:2024fud}. The PDFs used in GPDs are also different; in this section we used the JHA21 PDFs at $N^3LO$ approximation~\cite{Blumlein:2006be, Blumlein:2021lmf}. With these explanations, we drew the form factors for $u$ and $d$ quarks in Fig.(\ref{fig:tfuBL}), and the form factor diagrams for the JHA21 PDFs , especially for the $F_1^{d}$, are more consistent with the experimental data than KKA10 PDFs~\cite{Khorramian:2009xz}. (This result can be obtained by comparing Figs.(\ref{fig:tfud1}) and (\ref{fig:tfuBL})). The parameters of these PDFs are shown in Tab. (\ref{tab:tabpdf}). We have obtained these coefficients by fitting data from~\cite{Qattan:2012zf,Cates:2011pz,Diehl:2013xca}. The diagrams in Fig.(~\ref{fig:tfpBL}) depict nucleon form factors of the nucleons as a function of $-t$, calculated using the JHA21 PDFs~\cite{Blumlein:2006be, Blumlein:2021lmf} and various Ansatsez. The $G_E^{p,n}$ and $G_M^{p,n}$ as a function of $-t$ are shown in Fig.(\ref{fig:GEM BL}). The analysis shows that, especially for the neutron's electric form factors, the JHA21 PDFs are more consistent with experimental data than the KKA10 PDFs. Eqs. (\ref{eq:GN}) were used to arrive at this conclusion. In each of the paper's figures, we have analyzed the parameters of VS24 Ansatz for the upper and lower limits. Furthermore, using the necessary Ansatz, we computed these form factors in $q^2=0$ and compared them with the form factors derived from the experimental data of Ref.~\cite{Hohler:1976ax}. Tab. (\ref{tab:t=0}) shows the results of the computation. In general, it is difficult to find constant coefficients to fit all the electric and magnetic form factors of quarks as well as nucleons with the form factors obtained from DIS experiments. But Figs. (\ref{fig:tfud1}--\ref{fig:GEM BL}) in our studies show that the coefficients introduced in the Tab. (\ref{tab:t=0}) are suitable values for VS24 Ansatz and JHA21 PDFs. In Sec. (\ref{sec:sec4}), we discussed gravitional form factors $A_{q}(q^2)$ and $B_{q}(q^2)$, which is another application of GPDs. By combining the different PDFs and VS24 Ansatz, we calculated these form factors using Eqs. (\ref{eq:33}--\ref{eq:36}) and then plotted them all in Fig. (\ref{fig:AqBq}). Considering that there is no experimental data for gravitional form factors, we can see that the behavior of all graphs in this figure is the same. Also, at $t=0$, using different PDF combinations with VS24 Ansatz to calculate the form factors, the value of $A_{u+d}(t)$ is almost the same, which is evident in Tab.(\ref{tab:t=1}). Also, the proton GFF $A_{u+d}$ as a function of $-t$ for the MMNS (Lattice QCD)~\cite{More:2021stk} is plotted in Fig. (\ref{fig:Au+d}). Measuring the electric radii of protons and neutrons is one of the significant uses of form factors; therefore, in Sec. (\ref{sec:sec5}), we talk about the radii of nucleons. The Eqs. (\ref{eq:r1,p}--\ref{eq:rPMG3}) are used to compute the Dirac mean squared radii of the nucleon using the VS24 model. Tab.(\ref{tab:tabR}) displays the outcomes of these computations for VS24 Ansatz and JHA21 PDFs. The parameters that were selected for the VS24 ansatz match the data used in Ref. \cite{Xiong:2019umf}. A comprehensive comparison was made between the results of this Ansatz and PDFs in the $N^{3}LO$ approximation. Given that we are working with a combination of PDFs and Ansatz, VS24 Ansatz was introduced in order to fit the experimental data. This Ansatz was then combined with other PDFs and JHA21 PDFs, and it was determined that this combination of VS24 Ansatz and JHA21 PDFs better matches the experimental data. The optimal combination among the combinations listed in the article was also achieved by using additional Ansatzes. Also, the results of two groups, GJLY~\cite{Guo:2023pqw} and MMNS (Lattice QCD)~\cite{More:2021stk}, are plotted. Both the proton and neutron form factors produced results that were satisfactory with this combination.

%

\end{document}